\documentclass{acm_proc_article-sp}
\usepackage[english]{babel}
\usepackage{array}
\usepackage{supertabular}
\usepackage{graphicx}
\usepackage{cite}
\usepackage{url} 
\usepackage{hyperref}
\usepackage{times}

\begin{document}
\title{Automated generation of web server fingerprints}
\numberofauthors{3}
\author{
\alignauthor
Theodore Book\\
\affaddr{Rice University}
\email{tbook@rice.edu}
\alignauthor
Martha Witick\\
\affaddr{Rice University}
\email{martha.witick@rice.edu}
\alignauthor
Dan S. Wallach\\
\affaddr{Rice University}
\email{dwallach@rice.edu}
}
\maketitle

\abstract{
In this paper, we demonstrate that it is possible to automatically generate fingerprints for various web server types using multifactor Bayesian inference on randomly selected servers on the Internet, without building an a priori catalog of server features or behaviors.  This makes it possible to conclusively study web server distribution without relying on reported (and variable) version strings.  We gather data by sending a collection of specialized requests to 110,000 live web servers.  Using only the server response codes, we then train an algorithm to successfully predict server types independently of the server version string.  In the process, we note several distinguishing features of current web infrastructure.}

\section{
Introduction}

One of the fundamental tactics used by both attackers and defenders is to understand what specific systems might be under attack or require defense. Individual point releases will have known software bugs for which off-the-shelf exploits may already be available.~\cite{garfinkel2011web} While web servers often need to know a browser's specific user agent, e.g., to deliver content that works around known bugs or missing features of a given web client, web clients have no particular reason to know the exact version of a web server. This leads to a common mitigation, where a web server administrator will deliberately obfuscate version strings.  When a server gives \texttt{commodore64-\allowbreak HTTPD/1.1} or \texttt{'; DROP TABLE server\-types; --} as its version string, this behavior is obvious, but it is difficult to measure the frequency at which such modifications occur, or the true types of servers with obfuscated names.

Historically, extensive work has been done on fingerprinting operating system kernels~\cite{beverly2004robust}, and traffic classification~\cite{karagiannis2005blinc}, often making use of probabilistic techniques~\cite{moore2005internet}, and on fingerprinting web browser configurations~\cite{yen2009browser}, among other areas.  While various commercial applications use known distinguishing features of different web servers to identify server types and versions~\cite{httprint}, surprisingly little academic work has been conducted on the topic of web server fingerprinting, and much of what has been done has focused on detecting or preventing fingerprinting~\cite{yang2010improving}.

In our research, rather than manually cataloging every possible quirk, we wish to take a machine learning approach, sending a variety of carefully crafted requests to large numbers of web servers on the Internet, and training our system to make subsequent identifications. We take advantage of the fact that many installations don't obfuscate their published version information, allowing us to train our models with remarkable accuracy.

\section{
Design and Methodology}

Our central design principle was to determine if variations in web server behavior made it possible to determine server type (e.g. Apache) and version (e.g. 2.2.3) without a manual analysis of individual pieces of software.  We undertook to do this by surveying a broad selection of running web servers and comparing their responses to various requests.  In the process, we gathered information on some behaviors and features that might be related to potential exploits either against either the servers themselves or against browsers accessing the servers.

\paragraph{The HTTP Protocol}

Knowledge of the HTTP protocol is necessary to understand web server behavior and variations. First documented with version 0.9 in 1991, the HTTP protocol is now standardized on version 1.1, released by the Internet Engineering Task Force in its RFC 2068 of January 1997~\cite{rfc2068}, and updated with RFC 2616 in June of 1999~\cite{rfc2616}.  While the basic format of the protocol is relatively straightforward, there is significant opportunity for different implementations of the protocol to behave differently, particularly in selecting among the 41 different response codes included in the protocol.  Furthermore, many request and response headers are optional, and their ordering is not specified by the protocol.  This means that different server implementations will have measurably different behavior that can be used for the purpose of classification, even when the question of incomplete and incorrect implementations is set aside.  These variations, described in part below, make it possible to identify server types and versions by the various ways the protocol is implemented.

In designing our experimental methodology, it was essential not to compromise or
damage any of the servers that we studied. Therefore we chose a
selection of correct, if somewhat unusual, HTTP requests, and analyzed
the responses.  We chose requests that would enable us to gather identifying information
about the servers --- not simply the basic HTTP version string that
nominally represents the server type and version, but also a set of
responses that should enable the identification of the server based on
its behavior, and not only on its reported type. In designing our
experimental methodology, we also sought to obtain aggregate data about large numbers
of servers, as opposed to a detailed understanding of the security
profile of a few selected sites, and so we chose an approach that
involved crawling a large number of sites and cataloging responses,
rather than a manual inspection of a few hosts.

\subsection{Bayesian Classification}
\label{bayes}

Bayes' rule is a basic principle of probability that allows one to calculate the probability of a given hypothesis given a certain datum, when one knows the correlation between the datum and the hypothesized conclusion in the universe being considered.  Consider the common application of spam filtering, where Bayesian classification has been found to be extremely effective, even in the face of dedicated attempts to defeat it~\cite{androutsopoulos2000experimental}.  If one considers the presence of a certain term in an e-mail as the datum in question, as well as the frequency of that term in both spam and non-spam e-mails, then Bayes' rule makes it possible to classify the probability that a given e-mail is spam based on the presence of that term.  Bayesian inference allows multiple observations of this type to be combined, potentially allowing for more accurate classification of the e-mail as spam or not.  Bayesian techniques have been used effectively for intrusion detection~\cite{kruegel2003bayesian}, data mining~\cite{fayyad1996advances}, and as an effective technique in fields well beyond the scope of computer science~\cite{gelman2004bayesian}.

Our application is somewhat more complex than spam filtering, as we need to classify servers as belonging to one of many categories of types, and not simply as a binary quantity (spam or not spam.)  In doing so, we use Bayes' rule to calculate the probability of a given server belonging to each server type, and choose the type with the highest probability.  (See Section~\ref{prediction}) This also enables us to associate a degree of certainty with our predictions, as we know the likelihood that our observed characteristics belong to our calculated server type~\cite{hanson1991bayesian}.

\subsection{
Selection of Requests}

In developing our experimental methodology, we selected 10 separate requests to send to
every surveyed server.  A summary of the requests and their purposes is shown in Table~\ref{requests}.  While we begin with a standard HTTP GET to provide a baseline, the other requests are intended to identify features that may have different (or missing) implementations across different web servers.  They were also chosen with the aim of receiving the same response regardless of site configuration.  Thus, the only files that we access directly are files that should be present on most servers: the root URL, the \texttt{robots.txt} file, and the \texttt{favicon.ico} file.  While these files may not be present on every server, their presence on most servers enables a broad comparison across sites independent of site content.

\begin{table}
\begin{tabular}{|p{1.33in}|p{2in}|}
\hline
 \textbf{Request} &
 \textbf{Rationale}\\\hline
 An ordinary get request against the root URL &
 This request provided a
{\textquotedblleft}baseline.{\textquotedblright} As the most ordinary
request that any server would receive, we expected it to be handled in
a straightforward manner with a 200 response code.\\\hline
 A partial get request of 50 bytes against the
root URL &
 This request allowed us to test the
server's implementation of the HTTP partial get
feature.\\\hline
 A conditional get request for pages modified
after a future date against the root URL &
 This request allowed us to test the
server's implementation of the HTTP conditional get
feature.\\\hline
 A head request against the root URL &
 This request allowed us to test the
server's implementation of the HTTP head
feature.\\\hline
 An options request &
 This request allowed us to test the
server's implementation of the HTTP options
feature.\\\hline
 A trace request against the root URL &
 This request allowed us to test the
server's implementation of the HTTP trace
feature.\\\hline
A request for the root URL as a CSS stylesheet &
 This request is similar to a request that might
be generated by a browser experiencing a cross site scripting attack.
A modern browser should watch the content type of the response,
refusing to interpret the page as CSS if it is labeled as HTML.\\\hline
 A request for robots.txt as a CSS stylesheet &
 Similar to the previous request, it allows us to
test mime type support on a different file type (text).\\\hline
 A request for a relative URL above the root directory &
 This request is invalid and should be rejected on
any server. If it were honored, it would give the client access to
the server's entire file system.\\\hline
 A request for the favicon &
 This request allows us to survey another content
type, again checking for proper MIME type support.\\\hline
\end{tabular}
\caption{Requests used in measuring server behavior}
\label{requests}
\end{table}

\subsection{Datasets}

We prepared three distinct data sets. The first consisted of responses
from the Alexa top 10,000 sites, the second consisted of responses from
the top 100,000, and the third consisted of responses from the last
10,000 sites in the Alexa top 1 million. These three data sets,
representing a total of 1.2 million HTTP requests and responses,
allowed us to abstract behavior for different server types, and to
compare the differences between the largest (and presumably most
carefully maintained) sites and other sites that are likely not as
actively maintained or uniquely configured.

\subsection{Prediction of Server Types}
\label{prediction}

Our work sought to provide a methodology to identify server types beyond
their self reported version strings. With this in mind, we collected the server response codes from the requests above, and used them to calculate a distinct fingerprint for each type of web server.  We could then match the responses of an unknown web server against the fingerprints that we had developed for various server types, enabling us to predict its type and version.

We associated the individual fingerprints with specific server types and versions by using the version string provided with the responses.  In doing so, we did not assume that the data in the version string was necessarily correct.  Indeed, our premise was that it is often changed to report something other than the correct server type and version.  However, we did assume that, as these changes are made by individual administrators acting independently, that there would not be a consistent incorrect server string for a given fingerprint.  In this way, the incorrect responses, even if they constituted a significant percentage of the sample, could be filtered out as noise, while the single largest reported version string for any given fingerprint would be, in fact, the correct one.

Our fingerprints took the form of a dataset listing the frequency of each response for each server type and request type.  We generated the fingerprint dataset by training on the raw data in our primary dataset, calculating the probability of a given server version given that server's response to our requests using a standard Bayesian methodology, as discussed in Section~\ref{bayes}:

\begin{center}
$P \left( response | server \right) = 
  \frac{ P \left( response \cap server \right) }{ P \left( server \right) } $
\end{center}

Then, for each server in the set being tested, we sought to predict the
probability of each server type. Assuming that the response to one
request was independent from the response to a different request, we
used Bayes' rule to combine the probabilities for each request for each server type as follows:

\begin{center}

$ P \left( server | \bigcap_{i=0}^{n} response_{i} \right) = $

$ P
 \frac{ 
       P \left(server \right) \prod P \left( response_{i} | server \right) }
       { 
       P \left(server \right) \prod P \left( response_{i} | server \right) + 
       P \left( \overline{server} \right) \prod P \left( response_{i} | \overline{server} \right) 
       } $
\end{center}

This gave us a probability for each server type given the total set of
responses any given site returned. We then selected the most probable
server type as our prediction, reporting both the server type and the
probability that we had generated.

\subsection{Limitations}
Some attempts at disguising a server's type and version may go beyond changing the version string.  For example, a program called ServerMask by Port 80 Software changes the format of headers, cookies, and file extensions to make Microsoft IIS resemble a different type of server.  Because our techniques do not use those particular markers to fingerprint the server, those changes should not affect our success.  However, software that manipulated the response codes could easily confuse our algorithm.  Indeed, it is possible that some of the unusual response codes that we detected came from software seeking to do exactly that.  Additionally, our methodology assumes that a single web address corresponds to a single server technology (which may or may not be distributed across multiple physical machines).  However, in cases where different requests are handled by different server technologies (as may happen, for example, when separate caching servers are used) our attempts at identifying a single server technology used on the site can, at best, identify only one of the technologies used.

\section{Results}

Having developed our techniques, we applied them to our data set, enabling us to validate our methods and discover several interesting behaviors of current web servers.

\subsection{
Distribution of Server Types}

In order to determine the effectiveness of our technique, it was necessary to test it on real world data.  In doing so, we gathered information on server type from raw version strings, processed it with our algorithms, and examined the results.

\subsubsection{Reported server types}

{
While the HTTP standard provides for the use of
{\textquotedblleft}Product Tokens{\textquotedblright} (section
3.8)
in the {\textquotedblleft}Server{\textquotedblright} field of the
response header (section 14.38)~\cite{rfc2616},
many individual sites choose to modify their headers to exclude this
information or replace it with irrelevant or incorrect data. Thus, we
found server versions such as \texttt{Nintendo}, \texttt{All your base are belong to us}, and \texttt{My Arse} in addition to more standard strings such as \texttt{Apache/\allowbreak 2.2.20 (Ubuntu)}. This
penchant for customization not only reflects the universal human desire
for self{}-expression, but an understandable desire to hide sensitive
information from potential attackers and competitors.  The reported server types which we observed are summarized in Table~\ref{servertypes}}.

\subsubsection{Calculated server types}
\begin{figure}
\centering
\setlength{\unitlength}{0.240900pt}
\ifx\plotpoint\undefined\newsavebox{\plotpoint}\fi
\sbox{\plotpoint}{\rule[-0.200pt]{0.400pt}{0.400pt}}%
\begin{picture}(1050,630)(0,0)
\sbox{\plotpoint}{\rule[-0.200pt]{0.400pt}{0.400pt}}%
\put(110.0,82.0){\rule[-0.200pt]{4.818pt}{0.400pt}}
\put(90,82){\makebox(0,0)[r]{0\%}}
\put(969.0,82.0){\rule[-0.200pt]{4.818pt}{0.400pt}}
\put(110.0,154.0){\rule[-0.200pt]{4.818pt}{0.400pt}}
\put(90,154){\makebox(0,0)[r]{5\%}}
\put(969.0,154.0){\rule[-0.200pt]{4.818pt}{0.400pt}}
\put(110.0,227.0){\rule[-0.200pt]{4.818pt}{0.400pt}}
\put(90,227){\makebox(0,0)[r]{10\%}}
\put(969.0,227.0){\rule[-0.200pt]{4.818pt}{0.400pt}}
\put(110.0,299.0){\rule[-0.200pt]{4.818pt}{0.400pt}}
\put(90,299){\makebox(0,0)[r]{15\%}}
\put(969.0,299.0){\rule[-0.200pt]{4.818pt}{0.400pt}}
\put(110.0,372.0){\rule[-0.200pt]{4.818pt}{0.400pt}}
\put(90,372){\makebox(0,0)[r]{20\%}}
\put(969.0,372.0){\rule[-0.200pt]{4.818pt}{0.400pt}}
\put(110.0,444.0){\rule[-0.200pt]{4.818pt}{0.400pt}}
\put(90,444){\makebox(0,0)[r]{25\%}}
\put(969.0,444.0){\rule[-0.200pt]{4.818pt}{0.400pt}}
\put(110.0,517.0){\rule[-0.200pt]{4.818pt}{0.400pt}}
\put(90,517){\makebox(0,0)[r]{30\%}}
\put(969.0,517.0){\rule[-0.200pt]{4.818pt}{0.400pt}}
\put(110.0,589.0){\rule[-0.200pt]{4.818pt}{0.400pt}}
\put(90,589){\makebox(0,0)[r]{35\%}}
\put(969.0,589.0){\rule[-0.200pt]{4.818pt}{0.400pt}}
\put(190.0,82.0){\rule[-0.200pt]{0.400pt}{4.818pt}}
\put(190,41){\rotatebox{-45}{\makebox(0,0)[l]{0-10\%}}}
\put(190.0,569.0){\rule[-0.200pt]{0.400pt}{4.818pt}}
\put(270.0,82.0){\rule[-0.200pt]{0.400pt}{4.818pt}}
\put(270,41){\rotatebox{-45}{\makebox(0,0)[l]{10-20\%}}}
\put(270.0,569.0){\rule[-0.200pt]{0.400pt}{4.818pt}}
\put(350.0,82.0){\rule[-0.200pt]{0.400pt}{4.818pt}}
\put(350,41){\rotatebox{-45}{\makebox(0,0)[l]{20-30\%}}}
\put(350.0,569.0){\rule[-0.200pt]{0.400pt}{4.818pt}}
\put(430.0,82.0){\rule[-0.200pt]{0.400pt}{4.818pt}}
\put(430,41){\rotatebox{-45}{\makebox(0,0)[l]{30-40\%}}}
\put(430.0,569.0){\rule[-0.200pt]{0.400pt}{4.818pt}}
\put(510.0,82.0){\rule[-0.200pt]{0.400pt}{4.818pt}}
\put(510,41){\rotatebox{-45}{\makebox(0,0)[l]{40-50\%}}}
\put(510.0,569.0){\rule[-0.200pt]{0.400pt}{4.818pt}}
\put(589.0,82.0){\rule[-0.200pt]{0.400pt}{4.818pt}}
\put(589,41){\rotatebox{-45}{\makebox(0,0)[l]{50-60\%}}}
\put(589.0,569.0){\rule[-0.200pt]{0.400pt}{4.818pt}}
\put(669.0,82.0){\rule[-0.200pt]{0.400pt}{4.818pt}}
\put(669,41){\rotatebox{-45}{\makebox(0,0)[l]{60-70\%}}}
\put(669.0,569.0){\rule[-0.200pt]{0.400pt}{4.818pt}}
\put(749.0,82.0){\rule[-0.200pt]{0.400pt}{4.818pt}}
\put(749,41){\rotatebox{-45}{\makebox(0,0)[l]{70-80\%}}}
\put(749.0,569.0){\rule[-0.200pt]{0.400pt}{4.818pt}}
\put(829.0,82.0){\rule[-0.200pt]{0.400pt}{4.818pt}}
\put(829,41){\rotatebox{-45}{\makebox(0,0)[l]{80-90\%}}}
\put(829.0,569.0){\rule[-0.200pt]{0.400pt}{4.818pt}}
\put(909.0,82.0){\rule[-0.200pt]{0.400pt}{4.818pt}}
\put(909,41){\rotatebox{-45}{\makebox(0,0)[l]{90-100\%}}}
\put(909.0,569.0){\rule[-0.200pt]{0.400pt}{4.818pt}}
\put(110.0,82.0){\rule[-0.200pt]{0.400pt}{122.136pt}}
\put(110.0,82.0){\rule[-0.200pt]{211.751pt}{0.400pt}}
\put(989.0,82.0){\rule[-0.200pt]{0.400pt}{122.136pt}}
\put(110.0,589.0){\rule[-0.200pt]{211.751pt}{0.400pt}}
\put(172,82){\rule{8.9133pt}{32.2806pt}}
\put(172.0,82.0){\rule[-0.200pt]{0.400pt}{32.040pt}}
\put(172.0,215.0){\rule[-0.200pt]{8.672pt}{0.400pt}}
\put(208.0,82.0){\rule[-0.200pt]{0.400pt}{32.040pt}}
\put(172.0,82.0){\rule[-0.200pt]{8.672pt}{0.400pt}}
\put(252,82){\rule{8.9133pt}{5.7816pt}}
\put(252.0,82.0){\rule[-0.200pt]{0.400pt}{5.541pt}}
\put(252.0,105.0){\rule[-0.200pt]{8.672pt}{0.400pt}}
\put(288.0,82.0){\rule[-0.200pt]{0.400pt}{5.541pt}}
\put(252.0,82.0){\rule[-0.200pt]{8.672pt}{0.400pt}}
\put(332,82){\rule{8.9133pt}{8.9133pt}}
\put(332.0,82.0){\rule[-0.200pt]{0.400pt}{8.672pt}}
\put(332.0,118.0){\rule[-0.200pt]{8.672pt}{0.400pt}}
\put(368.0,82.0){\rule[-0.200pt]{0.400pt}{8.672pt}}
\put(332.0,82.0){\rule[-0.200pt]{8.672pt}{0.400pt}}
\put(412,82){\rule{8.9133pt}{4.818pt}}
\put(412.0,82.0){\rule[-0.200pt]{0.400pt}{4.577pt}}
\put(412.0,101.0){\rule[-0.200pt]{8.672pt}{0.400pt}}
\put(448.0,82.0){\rule[-0.200pt]{0.400pt}{4.577pt}}
\put(412.0,82.0){\rule[-0.200pt]{8.672pt}{0.400pt}}
\put(492,82){\rule{8.9133pt}{5.2998pt}}
\put(492.0,82.0){\rule[-0.200pt]{0.400pt}{5.059pt}}
\put(492.0,103.0){\rule[-0.200pt]{8.672pt}{0.400pt}}
\put(528.0,82.0){\rule[-0.200pt]{0.400pt}{5.059pt}}
\put(492.0,82.0){\rule[-0.200pt]{8.672pt}{0.400pt}}
\put(571,82){\rule{8.9133pt}{36.6168pt}}
\put(571.0,82.0){\rule[-0.200pt]{0.400pt}{36.376pt}}
\put(571.0,233.0){\rule[-0.200pt]{8.672pt}{0.400pt}}
\put(607.0,82.0){\rule[-0.200pt]{0.400pt}{36.376pt}}
\put(571.0,82.0){\rule[-0.200pt]{8.672pt}{0.400pt}}
\put(651,82){\rule{8.9133pt}{50.589pt}}
\put(651.0,82.0){\rule[-0.200pt]{0.400pt}{50.348pt}}
\put(651.0,291.0){\rule[-0.200pt]{8.672pt}{0.400pt}}
\put(687.0,82.0){\rule[-0.200pt]{0.400pt}{50.348pt}}
\put(651.0,82.0){\rule[-0.200pt]{8.672pt}{0.400pt}}
\put(731,82){\rule{8.9133pt}{115.391pt}}
\put(731.0,82.0){\rule[-0.200pt]{0.400pt}{115.150pt}}
\put(731.0,560.0){\rule[-0.200pt]{8.672pt}{0.400pt}}
\put(767.0,82.0){\rule[-0.200pt]{0.400pt}{115.150pt}}
\put(731.0,82.0){\rule[-0.200pt]{8.672pt}{0.400pt}}
\put(811,82){\rule{8.9133pt}{15.8994pt}}
\put(811.0,82.0){\rule[-0.200pt]{0.400pt}{15.658pt}}
\put(811.0,147.0){\rule[-0.200pt]{8.672pt}{0.400pt}}
\put(847.0,82.0){\rule[-0.200pt]{0.400pt}{15.658pt}}
\put(811.0,82.0){\rule[-0.200pt]{8.672pt}{0.400pt}}
\put(891,82){\rule{8.9133pt}{75.8835pt}}
\put(891.0,82.0){\rule[-0.200pt]{0.400pt}{75.643pt}}
\put(891.0,396.0){\rule[-0.200pt]{8.672pt}{0.400pt}}
\put(927.0,82.0){\rule[-0.200pt]{0.400pt}{75.643pt}}
\put(891.0,82.0){\rule[-0.200pt]{8.672pt}{0.400pt}}
\put(110.0,82.0){\rule[-0.200pt]{0.400pt}{122.136pt}}
\put(110.0,82.0){\rule[-0.200pt]{211.751pt}{0.400pt}}
\put(989.0,82.0){\rule[-0.200pt]{0.400pt}{122.136pt}}
\put(110.0,589.0){\rule[-0.200pt]{211.751pt}{0.400pt}}
\end{picture}

\caption{Confidence levels for servers whose reported type differed from the calculated one. (top 10k servers based on training data from top 100k)}
\label{confidence}
\end{figure}
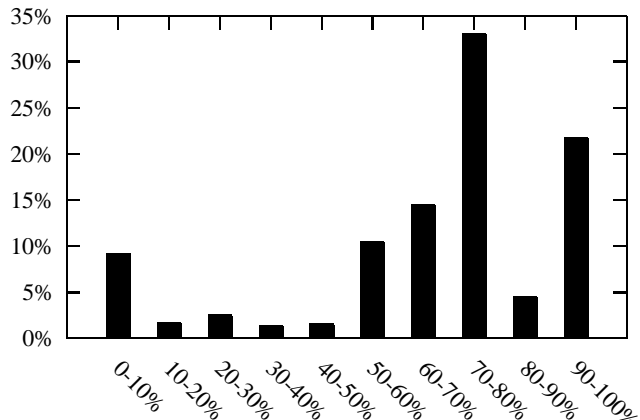

{
In order to correct for this behavior, we developed an algorithm
(described above) that used multifactor bayesian analysis to predict
the actual server type based on responses. Our algorithm produced
different results from the server type returned by the site for
approximately 38\% of sites. Figure~\ref{confidence} shows the confidence
levels of our predictions when our algorithm was trained on the top
100,000 sites, and run on the top 10,000 sites.}

While for a number of servers, we were able to predict the server version with a very high confidence level, in other cases, it is not immediately clear whether the server was reporting incorrect data, or whether our algorithm incorrectly predicted the type of the server.  As it was infeasible to contact the administrators of all servers in the sample to inquire as to the true server version, we relied upon the confidence levels produced by the Bayesian analysis.  Even beyond this, however, some circumstantial evidence does suggest that our algorithm correctly predicted the results in many cases.

Apart from the fact that our analysis produced the same results as the
server version string in 62\% of cases, some of the divergent results
also suggest correct behavior. For example, server versions which are
derived from Apache, such as IBM HTTP Server and Apache Coyote, were
recognized as Apache, even though they did not figure into the training
data for Apache. Most sites with unusual and obviously false version
strings were recognized as belonging to one of the major server types.

A more interesting question concerns the cases where servers with
apparently correct version strings were recognized as servers of a
different type. Are these cases of web site administrators seeking to
intentionally obfuscate their server selection, or cases of erroneous
behavior on the part of our software? As seen above, we were able to
generate a confidence estimate for each of our predictions, and many of
these results fell in the 75\% confidence range, suggesting a
likelihood that we were frequently correct, and the reported version strings represented an intentional effort at obfuscation.

\subsubsection{
Sites reporting multiple server types}

One phenomenon that affected our analysis was the case of a
number of sites that reported different server versions to different
requests. Within the top 100,000 websites, we found that 6 servers
gave 10 different version strings to our 10 requests, 84 gave 5 or
more, and an astounding 24,254 sites gave at least two different server
versions. One example is Verizon.com, which alternately reported
itself to be running Apache, Microsoft-IIS, Oracle-iPlanet-Web-Server
(the successor to Netscape Enterprise Server), AkamaiGHost, or returned
no version string at all!

While we have no direct evidence explaining this phenomenon, our data
does suggest several possible conclusions. For example, a number of
servers reported no server version at all, until they were asked for
their favicon.ico, at which time they reported that they were running
Microsoft-IIS. This suggested to us that a bug or configuration error
in Microsoft-IIS revealed the server version on that particular request
when an operator had desired for it to be hidden.

On other sites, it would appear that our requests were, indeed, being
handled by a variety of servers of different types. It would seem
that some common requests may have gone to a server handling static
content, while the more unusual requests got forwarded to a different
server which, in turn, returned a different server string. The
caching software used by Akamai Global Host, in particular, seemed to
produce this phenomenon.

It is conceivable that some web servers were configured to return different version strings to different requests, perhaps in an effort to confuse observers. However, we have no conclusive evidence to support this hypothesis.

\subsubsection{Server Types}

After running our analysis, we were able to classify the most common web servers found in our data set. The results are shown in Table~\ref{servertypes}. We give both the reported server types and the calculated server types for comparison.  As can be seen, the 15\% of servers which did not report a version have all been classified into one of the major versions.  

Many servers are re-classified as Apache, bringing the relatively low number of Apache servers found in the top 10,000 sites up to a level more typical of the internet as a whole.  Because we assign all servers to the most likely category, even when our confidence in our estimates are low, this may result in various obscure server versions being classified as Apache with a low confidence rating.  As the most common server type, Apache becomes the ``best guess'' when no good classification is possible.  

Additionally, the number of servers reporting Microsoft-IIS can be seen to have increased.  This is consistent with the known existence of software designed to obscure Microsoft-IIS servers.  More interesting is the virtual disappearance of Microsoft-HTTPAPI servers.  The Microsoft HTTP API is designed to let programs written in C service HTTP requests for specific URLs.  It is a relatively low-level API, exposing basic HTTP functionality to the programmer.  While there is no definitive way of knowing whether our re-classification of Microsoft HTTP API version strings was correct or not, it is worth noting that Microsoft HTTP API is not so much a server type as it is an interface for various programs to function as a web server.  Thus, it seems likely that there is limited commonality between the way different programs making use of this API handled our requests.  It may be the case that, because each program using this API is effectively a unique server type, that our algorithm was unable to effectively classify them.

Of the three smaller server versions, AkamiaGHost, cloudflare-nginx, and LiteSpeed, two are not server types at all, but content delivery networks.  CloudFlare and Akamai specialize in hosting static content in locations close to users, enabling faster page retrieval.  For this reason, it is not surprising that their servers should be re-classified as one of the common types --- their server version strings represent a delivery network and not a version type.  LiteSpeed, on the other hand, is a proprietary server technology.  While it advertises itself as being ``completely Apache interchangeable,''\cite{litespeed} LiteSpeed does not advertise itself as being an Apache derivative.  The general classification of LiteSpeed servers as Apache may reflect similarities in the behavior of the two servers as well as LiteSpeed's small market share.
\begin{table*}
\center
\tablehead{}
\begin{supertabular}{|m{1.2212598in}|r|r|>{\raggedleft}m{0.7in}|r|}
\hline
 \textbf{Server Type} &
 \textbf{Top 100k (Raw)} &
 \textbf{Top 10k (Raw)} &
 \textbf{Top 10k (Corrected)} &
 \textbf{Bottom  10k (Raw)}\\\hline
 Apache &
 40.7\% &
 29.6\% &
 59.6\% &
 52.8\%\\
 Nginx &
 12.8\% &
 16.1\% &
 17.6\% &
 8.2\%\\
 Microsoft{}-IIS &
 12.7\% &
 9.5\% &
 12.7\% &
 12.9\%\\
 Not Reported &
 12.3\% &
 15.5\% &
 None &
 9.1\%\\
 Microsoft{}-HTTPAPI &
 6.3\% &
 5.2\% &
 0.1\% &
 5.3\%\\
 AkamaiGHost &
 2.6\% &
 6.4\% &
 None &
 0.2\%\\
 Cloudflare{}-Nginx &
 1.5\% &
 0.9\% &
None &
 0.9\%\\
 LiteSpeed &
 1.5\% &
 0.8\% &
 None &
 1.2\%
\\\hline
\end{supertabular}
\caption{Prevalence of server types based on raw version strings and calculated values.}
\label{servertypes}
\end{table*}

\subsubsection{
Apache Versions}

In addition to the server type, the software version is an interesting topic of study.  It provides information as to how up-to-date a server is, an interesting question both for study and for potential attackers.  For this reason, we set out to see what information we could extract from our dataset regarding server versions.

Most servers in our dataset did not disclose their exact version, but a sufficient number provided the information to make some analysis possible. Because Apache servers possessed the largest market share --- and hence, the most complete data --- we chose to study their version strings in order to get a better understanding of the server population in active use. Although we have not studied other server types in the same detail, we expect these results to be similar to Apache derivatives and that the results can be generalized to other servers.

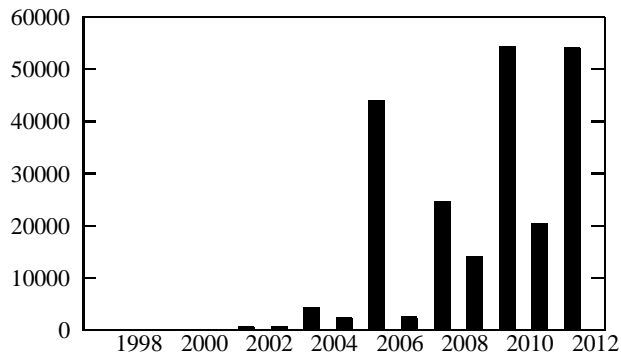
\begin{figure}
\centering
\setlength{\unitlength}{0.240900pt}
\ifx\plotpoint\undefined\newsavebox{\plotpoint}\fi
\sbox{\plotpoint}{\rule[-0.200pt]{0.400pt}{0.400pt}}%
\begin{picture}(1050,630)(0,0)
\sbox{\plotpoint}{\rule[-0.200pt]{0.400pt}{0.400pt}}%
\put(170.0,97.0){\rule[-0.200pt]{4.818pt}{0.400pt}}
\put(150,97){\makebox(0,0)[r]{ 0}}
\put(969.0,97.0){\rule[-0.200pt]{4.818pt}{0.400pt}}
\put(170.0,179.0){\rule[-0.200pt]{4.818pt}{0.400pt}}
\put(150,179){\makebox(0,0)[r]{ 10000}}
\put(969.0,179.0){\rule[-0.200pt]{4.818pt}{0.400pt}}
\put(170.0,261.0){\rule[-0.200pt]{4.818pt}{0.400pt}}
\put(150,261){\makebox(0,0)[r]{ 20000}}
\put(969.0,261.0){\rule[-0.200pt]{4.818pt}{0.400pt}}
\put(170.0,343.0){\rule[-0.200pt]{4.818pt}{0.400pt}}
\put(150,343){\makebox(0,0)[r]{ 30000}}
\put(969.0,343.0){\rule[-0.200pt]{4.818pt}{0.400pt}}
\put(170.0,425.0){\rule[-0.200pt]{4.818pt}{0.400pt}}
\put(150,425){\makebox(0,0)[r]{ 40000}}
\put(969.0,425.0){\rule[-0.200pt]{4.818pt}{0.400pt}}
\put(170.0,507.0){\rule[-0.200pt]{4.818pt}{0.400pt}}
\put(150,507){\makebox(0,0)[r]{ 50000}}
\put(969.0,507.0){\rule[-0.200pt]{4.818pt}{0.400pt}}
\put(170.0,589.0){\rule[-0.200pt]{4.818pt}{0.400pt}}
\put(150,589){\makebox(0,0)[r]{ 60000}}
\put(969.0,589.0){\rule[-0.200pt]{4.818pt}{0.400pt}}
\put(221,97){\usebox{\plotpoint}}
\put(221,589){\usebox{\plotpoint}}
\put(221,77){\makebox(0,0)[l]{1998}}
\put(272,97){\usebox{\plotpoint}}
\put(272,589){\usebox{\plotpoint}}
\put(324,97){\usebox{\plotpoint}}
\put(324,589){\usebox{\plotpoint}}
\put(324,77){\makebox(0,0)[l]{2000}}
\put(375,97){\usebox{\plotpoint}}
\put(375,589){\usebox{\plotpoint}}
\put(426,97){\usebox{\plotpoint}}
\put(426,589){\usebox{\plotpoint}}
\put(426,77){\makebox(0,0)[l]{2002}}
\put(477,97){\usebox{\plotpoint}}
\put(477,589){\usebox{\plotpoint}}
\put(528,97){\usebox{\plotpoint}}
\put(528,589){\usebox{\plotpoint}}
\put(528,77){\makebox(0,0)[l]{2004}}
\put(580,97){\usebox{\plotpoint}}
\put(580,589){\usebox{\plotpoint}}
\put(631,97){\usebox{\plotpoint}}
\put(631,589){\usebox{\plotpoint}}
\put(631,77){\makebox(0,0)[l]{2006}}
\put(682,97){\usebox{\plotpoint}}
\put(682,589){\usebox{\plotpoint}}
\put(733,97){\usebox{\plotpoint}}
\put(733,589){\usebox{\plotpoint}}
\put(733,77){\makebox(0,0)[l]{2008}}
\put(784,97){\usebox{\plotpoint}}
\put(784,589){\usebox{\plotpoint}}
\put(835,97){\usebox{\plotpoint}}
\put(835,589){\usebox{\plotpoint}}
\put(835,77){\makebox(0,0)[l]{2010}}
\put(887,97){\usebox{\plotpoint}}
\put(887,589){\usebox{\plotpoint}}
\put(938,97){\usebox{\plotpoint}}
\put(938,589){\usebox{\plotpoint}}
\put(938,77){\makebox(0,0)[l]{2012}}
\put(170.0,97.0){\rule[-0.200pt]{0.400pt}{118.523pt}}
\put(170.0,97.0){\rule[-0.200pt]{197.297pt}{0.400pt}}
\put(989.0,97.0){\rule[-0.200pt]{0.400pt}{118.523pt}}
\put(170.0,589.0){\rule[-0.200pt]{197.297pt}{0.400pt}}
\put(210,97){\rule{5.7816pt}{0.2409pt}}
\put(210,97){\usebox{\plotpoint}}
\put(210.0,97.0){\rule[-0.200pt]{5.541pt}{0.400pt}}
\put(210.0,97.0){\rule[-0.200pt]{5.541pt}{0.400pt}}
\put(261,97){\rule{5.7816pt}{0.2409pt}}
\put(261,97){\usebox{\plotpoint}}
\put(261.0,97.0){\rule[-0.200pt]{5.541pt}{0.400pt}}
\put(261.0,97.0){\rule[-0.200pt]{5.541pt}{0.400pt}}
\put(312,97){\rule{5.7816pt}{0.2409pt}}
\put(312,97){\usebox{\plotpoint}}
\put(312.0,97.0){\rule[-0.200pt]{5.541pt}{0.400pt}}
\put(312.0,97.0){\rule[-0.200pt]{5.541pt}{0.400pt}}
\put(363,97){\rule{5.7816pt}{0.2409pt}}
\put(363,97){\usebox{\plotpoint}}
\put(363.0,97.0){\rule[-0.200pt]{5.541pt}{0.400pt}}
\put(363.0,97.0){\rule[-0.200pt]{5.541pt}{0.400pt}}
\put(414,97){\rule{5.7816pt}{1.4454pt}}
\put(414.0,97.0){\rule[-0.200pt]{0.400pt}{1.204pt}}
\put(414.0,102.0){\rule[-0.200pt]{5.541pt}{0.400pt}}
\put(437.0,97.0){\rule[-0.200pt]{0.400pt}{1.204pt}}
\put(414.0,97.0){\rule[-0.200pt]{5.541pt}{0.400pt}}
\put(466,97){\rule{5.7816pt}{1.4454pt}}
\put(466.0,97.0){\rule[-0.200pt]{0.400pt}{1.204pt}}
\put(466.0,102.0){\rule[-0.200pt]{5.541pt}{0.400pt}}
\put(489.0,97.0){\rule[-0.200pt]{0.400pt}{1.204pt}}
\put(466.0,97.0){\rule[-0.200pt]{5.541pt}{0.400pt}}
\put(517,97){\rule{5.7816pt}{8.4315pt}}
\put(517.0,97.0){\rule[-0.200pt]{0.400pt}{8.191pt}}
\put(517.0,131.0){\rule[-0.200pt]{5.541pt}{0.400pt}}
\put(540.0,97.0){\rule[-0.200pt]{0.400pt}{8.191pt}}
\put(517.0,97.0){\rule[-0.200pt]{5.541pt}{0.400pt}}
\put(568,97){\rule{5.7816pt}{4.818pt}}
\put(568.0,97.0){\rule[-0.200pt]{0.400pt}{4.577pt}}
\put(568.0,116.0){\rule[-0.200pt]{5.541pt}{0.400pt}}
\put(591.0,97.0){\rule[-0.200pt]{0.400pt}{4.577pt}}
\put(568.0,97.0){\rule[-0.200pt]{5.541pt}{0.400pt}}
\put(619,97){\rule{5.7816pt}{86.9649pt}}
\put(619.0,97.0){\rule[-0.200pt]{0.400pt}{86.724pt}}
\put(619.0,457.0){\rule[-0.200pt]{5.541pt}{0.400pt}}
\put(642.0,97.0){\rule[-0.200pt]{0.400pt}{86.724pt}}
\put(619.0,97.0){\rule[-0.200pt]{5.541pt}{0.400pt}}
\put(670,97){\rule{5.7816pt}{5.0589pt}}
\put(670.0,97.0){\rule[-0.200pt]{0.400pt}{4.818pt}}
\put(670.0,117.0){\rule[-0.200pt]{5.541pt}{0.400pt}}
\put(693.0,97.0){\rule[-0.200pt]{0.400pt}{4.818pt}}
\put(670.0,97.0){\rule[-0.200pt]{5.541pt}{0.400pt}}
\put(722,97){\rule{5.7816pt}{48.6618pt}}
\put(722.0,97.0){\rule[-0.200pt]{0.400pt}{48.421pt}}
\put(722.0,298.0){\rule[-0.200pt]{5.541pt}{0.400pt}}
\put(745.0,97.0){\rule[-0.200pt]{0.400pt}{48.421pt}}
\put(722.0,97.0){\rule[-0.200pt]{5.541pt}{0.400pt}}
\put(773,97){\rule{5.7816pt}{27.7035pt}}
\put(773.0,97.0){\rule[-0.200pt]{0.400pt}{27.463pt}}
\put(773.0,211.0){\rule[-0.200pt]{5.541pt}{0.400pt}}
\put(796.0,97.0){\rule[-0.200pt]{0.400pt}{27.463pt}}
\put(773.0,97.0){\rule[-0.200pt]{5.541pt}{0.400pt}}
\put(824,97){\rule{5.7816pt}{107.441pt}}
\put(824.0,97.0){\rule[-0.200pt]{0.400pt}{107.200pt}}
\put(824.0,542.0){\rule[-0.200pt]{5.541pt}{0.400pt}}
\put(847.0,97.0){\rule[-0.200pt]{0.400pt}{107.200pt}}
\put(824.0,97.0){\rule[-0.200pt]{5.541pt}{0.400pt}}
\put(875,97){\rule{5.7816pt}{40.4712pt}}
\put(875.0,97.0){\rule[-0.200pt]{0.400pt}{40.230pt}}
\put(875.0,264.0){\rule[-0.200pt]{5.541pt}{0.400pt}}
\put(898.0,97.0){\rule[-0.200pt]{0.400pt}{40.230pt}}
\put(875.0,97.0){\rule[-0.200pt]{5.541pt}{0.400pt}}
\put(926,97){\rule{5.7816pt}{106.96pt}}
\put(926.0,97.0){\rule[-0.200pt]{0.400pt}{106.719pt}}
\put(926.0,540.0){\rule[-0.200pt]{5.541pt}{0.400pt}}
\put(949.0,97.0){\rule[-0.200pt]{0.400pt}{106.719pt}}
\put(926.0,97.0){\rule[-0.200pt]{5.541pt}{0.400pt}}
\put(170.0,97.0){\rule[-0.200pt]{0.400pt}{118.523pt}}
\put(170.0,97.0){\rule[-0.200pt]{197.297pt}{0.400pt}}
\put(989.0,97.0){\rule[-0.200pt]{0.400pt}{118.523pt}}
\put(170.0,589.0){\rule[-0.200pt]{197.297pt}{0.400pt}}
\end{picture}

\caption{Release year for Apache software currently in use}
\label{apacheage}
\end{figure}

Our first step in studying the distribution of Apache servers was simply
to examine the version strings returned by the servers themselves. As
is shown in Appendix B, the version numbers of Apache servers in active
use vary widely. While many are recent versions such as 2.2.22 and
2.2.23 (both 2012 releases) some servers use Apache versions dating
from the late 1990s and early 2000s. Indeed, more than 10,000
responses of the 963,000 received from the top 100,000 sites reported
some version of Apache 1.3, which reached its end of life in 2010,
although patches are still produced.

Server version strings, like server types, are subject to the caprice of
site maintainers, who have an incentive to report incorrect information
in an attempt at security by obscurity. Not only did slightly more than
half of the servers reporting themselves as running Apache give no
version details at all, but in the remainder, a few obviously false
version strings such as {\textquotedblleft}4.0.4{\textquotedblright}
and {\textquotedblleft}6.6.6{\textquotedblright} were returned.
However, it is unclear how many of the plausible version strings are
in fact false.

Using the same probabilistic techniques that we used to predict server
types, we attempted to predict server versions.  First, we took all of the servers that reported their server type as Apache, and trained our algorithm based on their reported versions from the top 100k.  We then applied the trained algorithm to the smaller data sets, seeking to predict the version for servers reporting Apache. Unfortunately, the
responses received from our queries did not meaningfully differentiate
Apache versions. It is possible that variations among Apache versions might enable a researcher to determine Apache versions sending different requests or by measuring different properties in HTTP responses, but we leave this to future work.

\subsubsection{Comparison with others' results}

One way to estimate the accuracy of our results is to compare them
against other sources. Netcraft gives results for
{\textquotedblleft}all domains{\textquotedblright} --- more than 625
million domains. They report the top servers as Apache with 57.2\%,
Microsoft with 16.5\%, nginx with 11.9\%, and Google with
3.4\%~\cite{netcraftsurvey}.
These results are largely consonant with our findings, although the
scale of the Netcraft survey is obviously much broader, and so the results can
not be directly compared. In a similar survey of 63.5 million servers,
Security Space reports Apache with 68.7\%, Microsoft with 14.89\%, and
Other (presumably including nginx) with
16.17\% of the market~\cite{sssurvey}.  While the results are largely consistent, we did not have access to the methodology for these two surveys, so we were unable to determine if they were naively relying on server version strings, or using some more sophisticated method of identifying server software.

\subsection{Server Responses}

In analyzing our data, we found that web servers returned 44 different
response codes, many of which do not form part of the official HTTP
specification. The table in Appendix A illustrates the range of
responses for our various requests. Some of these responses are
particularly noteworthy. For example, when we request text and HTML
documents as CSS, only a relatively small percentage of servers return
code 406, indicating that the document can not be returned given the
parameters sent by the client.  Interestingly enough, the number of servers returning the response varied depending on the request.  When we asked for a text file as CSS, 9,338 servers gave us a 406 error, while only 3,229 gave us that error when we asked for an HTML document as CSS.  Much more common is code 200, sending
the document despite the unusual request. While this response is
perfectly compliant with the HTTP standard, which provides for the
return of documents in a different type if the requested type is not
available, it may be a less wise response from a security perspective,
as the request suggests a cross site scripting attack is in progress,
and the attacker is looking for access to user data from that page.

Most servers correctly returned an error (400 --- bad request, 403 --- not
authorized, or 404 --- not found) to our request for a relative URL.
However, the 4,759 servers that returned a 200 code were not mostly
servers exposing a security vulnerability. Instead, they were mostly
configured to return a standard page for any malformed request - not a
strict interpretation of the HTTP standard, but not an immediate
security risk. For some reason, a 405 (not allowed) response was
returned by nearly half the servers for our trace request, rounding out
the significant anomalies found in the version codes.

\paragraph{Unsupported Features}

A number of the features used in our requests seemed simply not to be
supported by a majority of web servers. For example, our conditional
get, which asked for a page only if it had been modified after a future
date, only received the expected 304 response from some 7,000 of the
97,000 responding servers. The remainder simply served up the page,
even though the condition had clearly not been met. The actual 501
{\textquotedblleft}Not Implemented{\textquotedblright} response
appeared relatively rarely, with 14,000 servers returning it for our
trace request, and 3,700 returning it for our options request.

\subsection{Character Encoding}

Not surprisingly, a wide variety of character sets were returned by our
servers, with \texttt{UTF-8} being the most common, followed by \texttt{ISO 8859-1} and
\texttt{windows-1251}. A small portion of servers were obviously
misconfigured, returning an empty string or a string such as
\texttt{\$conf\_pass-{\textgreater}charset.}
We did not develop tests to see if the charsets for the remaining
servers matched the actual encoding of the content, but it seems
reasonable to assume that additional misconfigurations exist in that
space, as well.

More significant is the fact that more than one third of servers (36,000 out of
our 100,000) simply did not return a character encoding
at all. While correct charset encoding is more important to user
experience than to security, several existing security vulnerabilities
relate to incorrect or missing character encoding, indicating this is a
risk on some deployed servers~\cite{watson2007web}.

\subsection{MIME Types}

Also interesting from a security perspective is the range of MIME types
received. While the responses to our standard GET request were mostly
the expected \texttt{text/\allowbreak html}, the responses to some our other requests had
more variety. Responses such as \texttt{application/\allowbreak java-\allowbreak archive} and \texttt{application/\allowbreak json} for an HTML document hinted at server
misconfigurations. Significantly, 0.1\% of servers returned no MIME
type at all, exposing themselves to exploits where an attacker crafts a
site to cross-load resources as a different MIME type, thus extracting
data from a confidential document.

\subsection{XSS Vulnerabilities}

One of the more curious MIME type handling behaviors that we observed was found in a number of servers that generate the MIME type for their response based not on the actual type of the file, but on the type requested by the client. Indeed, 208 sites in the top 100k, mostly reporting Nginx as their server type, exhibited this behavior. This is a potential vulnerability, as it opens clients to cross site scripting attacks~\cite{huang2010protecting}.

In a cross origin CSS attack, an HTML page with confidential information is loaded by a script running from an attacker's site.  If certain short CSS sequences have been injected into the page, the attacker is able to read portions of the secure page.  Most browsers, including Chrome, Safari, and Firefox, prevent this by refusing to load files as CSS resources cross origin if they present a conflicting MIME type. However, when asked for a HTML page as CSS, these sites, including sites like kickstarter.com and causes.com which handle sensitive data, report that the HTML page is, in fact, CSS, potentially enabling this sort of attack against their users.  While browsers could protect themselves against these attacks by determining a document's content type through analysis, this behavior represents a significant weakness~\cite{barth2009secure}.

\section{Future Work}

Our results show that it is possible to gather significant data about a
web server and its vulnerabilities based on a few carefully selected
HTTP requests. However, there is much room available for further work
in refining this process. In particular, additional work is needed to
develop a library of requests that are particularly effective at
gathering distinguishing information about a server and its security
vulnerabilities. Our requests were designed on the basis of informed
judgment, but systematic tests with a larger collection of possible
requests would likely result in a collection that allowed greater
accuracy in predicting server type and in detecting vulnerabilities.

Additionally, our analysis focused primarily on the response codes and
MIME types returned from the server. Further research could
investigate other details of the server's response,
perhaps finding variations that enable better classification and
analysis of the servers studied. Furthermore, ongoing research could
survey a broader selection of servers, both less visited servers
hosting sites for the general public and the embedded servers found in
so many devices.

While our work focused on web servers, and much similar work has been done on general network traffic, there are a broad range of additional Internet technologies where probabilistic techniques could be used to yield a better understanding of software distribution, maintenance, and behavior.  This presents many interesting areas for ongoing study.

\section{Conclusions}

We have demonstrated that it is possible to predict server type based on server responses, even without a prior analysis of server behavior.  This enables a more accurate classification of web server types than would be possible using server version strings alone.  It also shows that attempts at security through obscurity by hiding the server version do not provide effective protection to vulnerable web servers.  While we were not able to effectively calculate server versions using the technique of analyzing response codes to individual requests, there is no fundamental reason why similar, but more precise, techniques might not also be able to differentiate among versions.

Additionally, even our broad survey, not focused on any particular site or vulnerability, was able to reveal a number of potentially serious configuration issues.  This is despite the fact that we focused only on the most visited websites, which would presumably be among the best maintained.  Among the misconfigurations observed were improper MIME type configurations that allow a document to be served with no MIME type or with an incorrect type based on the client's request.  Dated server software was also found to be widely in use.  The maintenance of a web site is, of course, the responsibility of the site's owner, but we hope that our research may provide a minor stimulus to raising the profile of the dangers caused by mis-configured server software.

\section{Acknowledgements}
...

\bibliographystyle{acm}
\bibliography{serversurvey}

\section{Appendix A: Responses from Servers}
Table~\ref{responses} gives a listing of all of the response strings returned by the top 100,000 servers to our various requests.  As can be seen, most of the responses cluster into a few expected categories, but there are large numbers of outliers, including many version codes not defined in the official standard.  Even some official codes occurred in surprising places.  For example, the IETF defines response code 418 as being an appropriate response from a teapot when instructed to brew coffee~\cite{rfc2324}.  As none of our requests were related to the preparation of hot beverages, it is safe to assume that this code was being used outside of its defined scope.

\begin{figure*}
\begin{flushleft}
\tablehead{}
\begin{supertabular}{|m{0.5in}|>{\raggedleft}m{0.49135986in} >{\raggedleft}m{0.49135986in} >{\raggedleft}m{0.49135986in} >{\raggedleft}m{0.49135986in} >{\raggedleft}m{0.49135986in} >{\raggedleft}m{0.49135986in} >{\raggedleft}m{0.49135986in} >{\raggedleft}m{0.49135986in} >{\raggedleft}m{0.49135986in} >{\raggedleft}m{0.49135986in} |}
\hline
 \textbf{Response Code} &
 \textbf{Condit-ional Get} &
 \textbf{Get FavIcon} &
 \textbf{Get Relative URL} &
 \textbf{Get} &
 \textbf{Head} &
 \textbf{Get HTML as CSS} &
 \textbf{Options} &
 \textbf{Partial Get} &
 \textbf{Trace} &
 \textbf{Get text as CSS}\tabularnewline
 \hline
  0 &	~	 &	~	 &	~	 &	~	 &	~	 &	~	 &	~	 &	~	 &	~	 &	 1\tabularnewline
	 1 &	~	 &	~	 &	~	 &	~	 &	~	 &	~	 &	~	 &	~	 &	~	 &	 1\tabularnewline
	 200 &	 88,689 &	 73,599 &	 4,759 &	 95,654 &	 94,059 &	 91,877 &	 80,192 &	 95,653 &	 27,728 &	 65,003\tabularnewline
	 203 &	~	 &	~	 &	~	 &	~	 &	 3 &	~	 &	~	 &	~	 &	~	 &	~	\tabularnewline
	 204 &	 3 &	 79 &	 6 &	 3 &	 11 &	 3 &	 10 &	 3 &	 2 &	 4\tabularnewline\hline
	 205 &	~	 &	 1 &	 3 &	~	 &	~	 &	~	 &	 4 &	~	 &	~	 &	 2\tabularnewline
	 300 &	~	 &	 3 &	~	 &	~	 &	~	 &	~	 &	~	 &	~	 &	~	 &	 2\tabularnewline
	 301 &	 2 &	 1 &	 1 &	 2 &	 4 &	 2 &	 3 &	 2 &	 3 &	 1\tabularnewline
	 302 &	 1 &	 3 &	~	 &	 1 &	 1 &	 3 &	~	 &	 1 &	~	 &	 3\tabularnewline
	 304 &	 6,985 &	 1 &	~	 &	~	 &	~	 &	~	 &	~	 &	~	 &	 1 &	~	\tabularnewline\hline
	 400 &	 69 &	 72 &	 66,759 &	 70 &	 148 &	 70 &	 385 &	 69 &	 289 &	 66\tabularnewline
	 401 &	 26 &	 26 &	 11 &	 26 &	 28 &	 24 &	 271 &	 26 &	 200 &	 25\tabularnewline
	 402 &	~	 &	~	 &	~	 &	~	 &	~	 &	~	 &	 2 &	~	 &	~	 &	~	\tabularnewline
	 403 &	 411 &	 268 &	 16,331 &	 410 &	 952 &	 407 &	 1,676 &	 409 &	 9,492 &	 405\tabularnewline
	 404 &	 357 &	 22,455 &	 6,616 &	 356 &	 540 &	 417 &	 1,221 &	 359 &	 1,508 &	 21,547\tabularnewline\hline
	 405 &	~	 &	~	 &	 3 &	~	 &	 243 &	~	 &	 7,768 &	~	 &	 40,968 &	 1\tabularnewline
	 406 &	 1 &	 5 &	 6 &	 1 &	 3 &	 3,229 &	 39 &	 1 &	 45 &	 9,338\tabularnewline
	 407 &	~	 &	~	 &	~	 &	~	 &	~	 &	~	 &	 2 &	~	 &	 2 &	~	\tabularnewline
	 408 &	 2 &	 2 &	 3 &	 2 &	 2 &	 3 &	 2 &	 2 &	 2 &	 2\tabularnewline
	 409 &	 1 &	 1 &	~	 &	 1 &	 1 &	 1 &	 1 &	 1 &	~	 &	 1\tabularnewline\hline
	 410 &	 1 &	 20 &	 34 &	 1 &	 1 &	 1 &	 1 &	 2 &	 3 &	 10\tabularnewline
	 411 &	~	 &	~	 &	~	 &	~	 &	~	 &	~	 &	 21 &	~	 &	~	 &	~	\tabularnewline
	 412 &	~	 &	~	 &	~	 &	~	 &	~	 &	~	 &	 1 &	~	 &	~	 &	~	\tabularnewline
	 413 &	~	 &	~	 &	~	 &	~	 &	~	 &	~	 &	~	 &	~	 &	 17 &	~	\tabularnewline
	 417 &	~	 &	~	 &	 25 &	~	 &	~	 &	~	 &	~	 &	~	 &	~	 &	~	\tabularnewline\hline
	 418 &	~	 &	~	 &	 1 &	~	 &	~	 &	~	 &	~	 &	~	 &	~	 &	~	\tabularnewline
	 420 &	~	 &	~	 &	~	 &	~	 &	~	 &	~	 &	 1 &	~	 &	~	 &	~	\tabularnewline
	 422 &	~	 &	~	 &	~	 &	~	 &	~	 &	~	 &	 11 &	~	 &	~	 &	~	\tabularnewline
	 429 &	 1 &	~	 &	~	 &	 1 &	 1 &	 1 &	~	 &	~	 &	~	 &	~	\tabularnewline
	 440 &	~	 &	~	 &	~	 &	~	 &	 1 &	~	 &	~	 &	~	 &	~	 &	~	\tabularnewline\hline
	 499 &	~	 &	~	 &	 1 &	~	 &	~	 &	~	 &	~	 &	~	 &	~	 &	~	\tabularnewline
	 500 &	 119 &	 146 &	 832 &	 124 &	 253 &	 533 &	 335 &	 122 &	 196 &	 193\tabularnewline
	 501 &	~	 &	 1 &	 8 &	~	 &	 31 &	 2 &	 3,663 &	~	 &	 14,002 &	~	\tabularnewline
	 502 &	 53 &	 34 &	 11 &	 53 &	 63 &	 56 &	 87 &	 52 &	 12 &	 44\tabularnewline
	 503 &	 88 &	 57 &	 85 &	 87 &	 192 &	 94 &	 158 &	 85 &	 76 &	 59\tabularnewline\hline
	 504 &	 10 &	 7 &	 11 &	 12 &	 11 &	 9 &	 10 &	 13 &	 7 &	 6\tabularnewline
	 508 &	 2 &	 1 &	 2 &	 1 &	 1 &	 4 &	 1 &	 1 &	 1 &	 1\tabularnewline
	 509 &	 1 &	 1 &	 6 &	 1 &	 1 &	 1 &	 1 &	 1 &	 1 &	 1\tabularnewline
	 550 &	~	 &	~	 &	~	 &	~	 &	 1 &	~	 &	~	 &	~	 &	 1 &	~	\tabularnewline
	 599 &	~	 &	~	 &	~	 &	~	 &	 1 &	~	 &	~	 &	~	 &	~	 &	~	\tabularnewline\hline
	 770 &	 1 &	 1 &	 1 &	 1 &	 1 &	 1 &	 1 &	 1 &	~	 &	 1\tabularnewline
	 801 &	~	 &	~	 &	 1 &	~	 &	~	 &	~	 &	 1 &	~	 &	 1 &	~	\tabularnewline
	 901 &	~	 &	~	 &	~	 &	~	 &	~	 &	~	 &	~	 &	~	 &	~	 &	 1\tabularnewline
	 999 &	~	 &	~	 &	~	 &	~	 &	~	 &	~	 &	 6 &	~	 &	 29 &	~	\tabularnewline\hline
	 \textbf{Totals} &	 96,823 &	 96,784 &	 95,516 &	 96,807 &	 96,553 &	 96,738 &	 95,874 &	 96,803 &	 94,586 &	 96,718\tabularnewline\hline

\end{supertabular}
\end{flushleft}
\label{responses}
\caption{Server response codes from top 100,000 servers by request}
\end{figure*}

\clearpage
\section{
Appendix B: Apache Versions}

{
The following versions of Apache were reported by servers in the Alexa
top 100k. A string such as 2.X.X indicates that the server only
reported Apache 2, and so on.}

\begin{flushleft}
\tabletail{
\hline
}
\tablehead{
\hline
 \textbf{Apache Version} &
 \textbf{No. of Samples Reported} &
 \textbf{Release Date}\\
 \hline
}
\begin{supertabular}{|m{1in}|>{\raggedleft}m{0.9in}|m{1in}|}

 	 Apache/1.3.3 &	 9 &	 9-Oct{}-98\\
 	Apache/1.3.9 &	 2 &	 19-Aug{}-99\\
	Apache/1.3.11 &	 5 &	 21-Jan{}-00\\
	 Apache/1.3.19 &	 8 &	 1-Mar{}-01\\
	 Apache/1.3.20 &	 40 &	 21-May{}-01\\
	 Apache/1.3.22 &	 10 &	 12-Oct{}-01\\
	 Apache/1.3.23 &	 26 &	 21-Jan{}-02\\
	 Apache/1.3.24 &	 27 &	 22-Mar{}-02\\
	 Apache/1.3.26 &	 177 &	 18-Jun{}-02\\
	 Apache/1.3.27 &	 265 &	 3-Oct{}-02\\
	 Apache/1.3.28 &	 40 &	 16-Jul{}-03\\
	 Apache/1.3.29 &	 264 &	 29-Oct{}-03\\
	 Apache/1.3.31 &	 180 &	 11-May{}-04\\
	 Apache/1.3.32 &	 8 &	 Not released\\
	 Apache/1.3.33 &	 559 &	 29-Oct{}-04\\
	 Apache/1.3.34 &	 509 &	 18-Oct{}-05\\
	 Apache/1.3.35 &	 19 &	 1-May{}-06\\
	 Apache/1.3.36 &	 77 &	 17-May{}-06\\
	 Apache/1.3.37 &	 1,223 &	 28-Jul{}-06\\
	 Apache/1.3.39 &	 271 &	 7-Sep{}-07\\
	 Apache/1.3.41 &	 2,964 &	 19-Jan{}-08\\
	 Apache/1.3.42 &	 3,618 &	 2-Feb-08 [EOL]\\
	 Apache/1.4.0 &	 10 &	 n/a\\
	 Apache/1.4.X &	 100 &	 n/a\\
	 Apache/1.9.0 &	 1 &	 n/a\\
	 Apache/2.0.4 &	 10 &	 n/a\\
	 Apache/2.0.6 &	 10 &	 n/a\\
	 Apache/2.0.29 &	 10 &	 Not released\\
	 Apache/2.0.35 &	 7 &	 5-Apr{}-02\\
	 Apache/2.0.40 &	 96 &	 9-Aug{}-02\\
	 Apache/2.0.43 &	 10 &	 3-Oct{}-02\\
	 Apache/2.0.44 &	 10 &	 20-Jan{}-03\\
	 Apache/2.0.45 &	 10 &	 1-Apr{}-03\\
	 Apache/2.0.46 &	 270 &	 28-May{}-03\\
	 Apache/2.0.47 &	 10 &	 9-Jul{}-03\\
	 Apache/2.0.48 &	 20 &	 29-Oct{}-03\\
	 Apache/2.0.49 &	 61 &	 19-Mar{}-04\\
	 Apache/2.0.50 &	 70 &	 30-Jun{}-04\\
	 Apache/2.0.51 &	 144 &	 15-Sep{}-04\\
	 Apache/2.0.52 &	 3,107 &	 28-Sep{}-04\\
	 Apache/2.0.53 &	 132 &	 7-Feb{}-05\\
	 Apache/2.0.54 &	 450 &	 17-Apr{}-05\\
	 Apache/2.0.55 &	 348 &	 16-Oct{}-05\\
	 Apache/2.0.58 &	 122 &	 1-May{}-06\\
	 Apache/2.0.59 &	 1,226 &	 28-Jul{}-06\\
	 Apache/2.0.61 &	 204 &	 7-Sep{}-07\\
	 Apache/2.0.63 &	 4,858 &	 19-Jan{}-08\\
	 Apache/2.0.64 &	 3,804 &	 19-Oct{}-10\\
	 Apache/2.0.X &	 133 &	 on or after March 10, 2000\\
	 Apache/2.1.X &	 20 &	 before December 1, 2005\\
	 Apache/2.2.0 &	 892 &	 1-Dec{}-05\\
	 Apache/2.2.1 &	 10 &	 Not released\\
 	 Apache/2.2.2 &	 214 &	 1-May{}-06\\
 	 Apache/2.2.3 &	 41,012 &	 28-Jul{}-06\\
	 Apache/2.2.4 &	 927 &	 9-Jan{}-07\\
	 Apache/2.2.6 &	 1008 &	 7-Sep{}-07\\
	 Apache/2.2.8 &	 3,443 &	 19-Jan{}-08\\
	 Apache/2.2.9 &	 8,635 &	 14-Jun{}-08\\
	 Apache/2.2.10 &	 1,401 &	 14-Oct{}-08\\
	 Apache/2.2.11 &	 3,270 &	 14-Dec{}-08\\
	 Apache/2.2.12 &	 1,126 &	 28-Jul{}-09\\
	 Apache/2.2.13 &	 819 &	 8-Aug{}-09\\
	 Apache/2.2.14 &	 12,018 &	 3-Oct{}-09\\
	 Apache/2.2.15 &	 15,230 &	 5-Mar{}-10\\
	 Apache/2.2.16 &	 18,947 &	 25-Jul{}-10\\
	 Apache/2.2.17 &	 12,667 &	 18-Oct{}-10\\
	 Apache/2.2.18 &	 299 &	 11-May{}-11\\
	 Apache/2.2.19 &	 4,108 &	 21-May{}-11\\
	 Apache/2.2.20 &	 4,010 &	 30-Aug{}-11\\
	 Apache/2.2.21 &	 11,910 &	 13-Sep{}-11\\
	 Apache/2.2.22 &	 39,446 &	 31-Jan{}-12\\
	 Apache/2.2.23 &	 13,660 &	 13-Sep{}-12\\
	 Apache/2.2.X &	 6,071 &	 on or after December 1, 2005\\
 	 Apache/2.3.5 &	 20 &	 26-Jan{}-10\\
	 Apache/2.3.6 &	 10 &	 17-Jun{}-10\\
	 Apache/2.3.8 &	 28 &	 31-Aug{}-10\\
	 Apache/2.3.11 &	 7 &	 7-Mar{}-11\\
	 Apache/2.3.14 &	 7 &	 9-Aug{}-11\\
	 Apache/2.3.16 &	 6 &	 20-Dec{}-11\\
	 Apache/2.4.0 &	 8 &	 Not released\\
	 Apache/2.4.1 &	 182 &	 17-Feb{}-12\\
	 Apache/2.4.2 &	 257 &	 17-Apr{}-12\\
	 Apache/2.4.3 &	 506 &	 21-Aug{}-12\\
	 Apache/2.4.X &	 40 &	 2012\\
	 Apache/2.X.X &	 10,891 &	 on or after March 10, 2000\\
 \hline
 \textbf{\# of samples reporting version:} &
 \textbf{238,508} &
~
\\
\end{supertabular}
\end{flushleft}

%

\end{document}